\documentclass[12pt,preprint]{aastex}

\usepackage{emulateapj5}
\usepackage{epsf}
\usepackage{graphics}

\newenvironment{inlinefigure}{
\def\@captype{figure}
\noindent\begin{minipage}{0.999\linewidth}\begin{center}}
{\end{center}\end{minipage}\smallskip}
\vspace{1.5cm}
\slugcomment{Accepted to the Astrophysical Journal Letters}

\shorttitle{The Universal Initial Galaxy Mass Function}
\shortauthors{CONSELICE}

\begin{document}

\title{The Formation of Low-Mass Cluster Galaxies and the Universal
Initial Galaxy Mass Function}

\author{Christopher J. Conselice}

\affil{California Institute of Technology, Mail Code 105-24, Pasadena
CA 91125}

\begin{abstract}

Clusters of galaxies have an observed over-density of low-luminosity systems in
comparison to the field, although it is not yet agreed whether this effect
is the result of initial galaxy mass functions that vary with environment
or galaxy evolutionary effects.  In this letter we argue that this over-density
is the result of low-mass systems with red colors 
that are over-populating the faint-end of the observed luminosity function in 
the nearby rich cluster Abell~0426. We
show that the luminosity function of Abell~0426 becomes steeper, 
from the field value $\alpha$ = -1.25$\pm$0.05 to $\alpha$=-1.44$\pm$0.04, 
due to a recently identified population of red low-mass cluster galaxies that 
are possibly the remnants of dynamical stripped high-mass systems.  
We further demonstrate, through simple models of 
stripping effects, how cluster luminosity functions can become 
artificially steep over time from the production of these low-mass cluster 
galaxies.

\end{abstract}


\section{Introduction}

Clusters of galaxies have long been observed to contain an 
over-density of low-luminosity galaxies relative to brighter systems in
comparison to the field 
(e.g., Binggeli, Sandage \& Tammann 1988; Ferguson \& Binggeli 1994).  This 
naturally includes observing a higher number of galaxies at the faint end of 
the luminosity function (LF), resulting in a steeper faint-end 
power-law and Schechter function slopes (e.g., Trentham 1998).  This has
led to the idea that the luminosity function of clusters is
fundamentally different than the field LF, and that the galaxy formation
process changes in different environments 
(e.g., Binggeli et al. 1988). Galaxy cluster LFs can be altered by galaxy
evolutionary effects, as we now known that the bright end of the 
LFs of groups and clusters are affected by merging events (e.g., Dressler et 
al. 1997; Jones et al. 2000).  Faint low-mass galaxies
are however usually spared a merging fate as 
their low-masses saves them from dynamical friction effects. A steeper LF 
can however be produced by creating new dwarfs or low-mass
galaxies from higher mass systems due to the removal of mass.   
The processes of creating low-luminosity 
galaxies from more luminous ones has not generally been considered in terms
of its effect on the observed LFs of clusters.

Recent observational results suggest that a significant fraction of 
low-mass cluster galaxies (LMCGs) were not part of the initial cluster 
luminosity or  mass function but were formed later possibly through the 
dynamical stripping of higher mass galaxies (e.g.,  
Gallagher, Conselice \& Wyse 2001; Conselice, Gallagher, \& Wyse 2002a,b). 
Evidence for this includes: kinematic information
suggesting that some dwarf ellipticals/spheroidals were 
accreted into clusters during the last few Gyrs (e.g., Conselice
et al. 2001; Drinkwater et al. 2001) and metal rich stellar populations in
low-mass galaxies 
(e.g., Adami et al. 2000; Rakos et al. 2001; Conselice et al. 2002b) 
suggesting that LMCGs are possibly
the remnants of stripped massive systems (Conselice et al. 2002b). 

Some low-mass cluster galaxies however have properties, such as low 
metallicity stellar populations, that suggest they originally
formed with the cluster (e.g., Conselice et al. 2001; Conselice et al. 2002b).
Low-mass cluster galaxies are therefore a combination of at least
two populations with possibly different evolutionary histories; one 
put into  place during cluster formation as predicted in CDM models 
(e.g., Nagamine et al. 2001a) 
and the other from evolved systems that formed after the cluster itself,  
possibly through tidal mechanisms such as harassment (e.g., Moore et al. 
1998).   This letter explores the implications of this later scenario for 
cluster luminosity 
functions, and how cluster luminosity functions could be derived from a 
universal initial galaxy
mass function. The cosmology H$_{0}$ = 70 km s$^{-1}$ Mpc$^{-1}$, 
$\Omega_{\lambda}$ = 0.7, and $\Omega_{\lambda}$ = 0.3, is
used throughout this letter.

\section{The Dependence of Color on Cluster Luminosity Functions}

The data for this letter are presented in detail in Conselice et al. 
(2002a,b) and consist of deep (4.4 ks) broad-band B and R 
photometry of the central
regions of the cluster Abell~0426 (Perseus) located at a distance of 77 Mpc.
This imaging was acquired with the WIYN 3.5~m telescope in good 
($\sim$0.7\arcsec) seeing.  See
Conselice et al. (2002a) for details of the observations and reduction 
procedures for this data. Conselice et al. 
(2002a,b) suggest that in the central $\sim 173$ arcmin$^2$ of Abell~0426 
the low-mass galaxy systems have a broad range of colors, which we 
arbitrarily separate into a red and blue population.  The separation of 
low-mass systems in Abell~0426
from background systems is discussed in detail in Conselice et al. (2002a).
To distinguish low-mass cluster members from background galaxies,
Conselice et al. 
(2002a) used morphological, color, size, surface brightness and magnitude
cuts.  Figure~1 shows the color-magnitude diagram of the galaxies
identified in the center of Abell~0426,  with $M_{B} < -11$, demonstrating
the broad color range of the LMCGs.

For discussion purposes, we arbitrarily define blue LMCGs as galaxies which are
$< 2 \sigma$ away, or bluer, than the color-magnitude relationship as
defined by the giant elliptical galaxy color scatter at M$_{B} = -17$ 
(e.g,. Conselice et al. 2002a).  LMCGs which are $>2\sigma$ redder than the 
CMR are denoted as  red systems.  This 2$\sigma$ separation is shown as the 
dashed line
in Figure~1. Conselice et al. (2002b) show that the red LMCGs cannot be 
accounted for as old low-mass systems within standard star-formation scenarios 
(e.g., Dekel \& Silk 1986; Nagamine et al. 2001a) due to their
high inferred metallicities, but the blue low-mass
cluster galaxies have properties that fit these early formation models. 
The blue LMCG color distribution can also be fit by a gaussian, while the red 
and total LMCG color distributions are not well fit by gaussians,  as they 
both have a broad distribution in color.

\begin{inlinefigure}
\begin{center}
\resizebox{\textwidth}{!}{\includegraphics{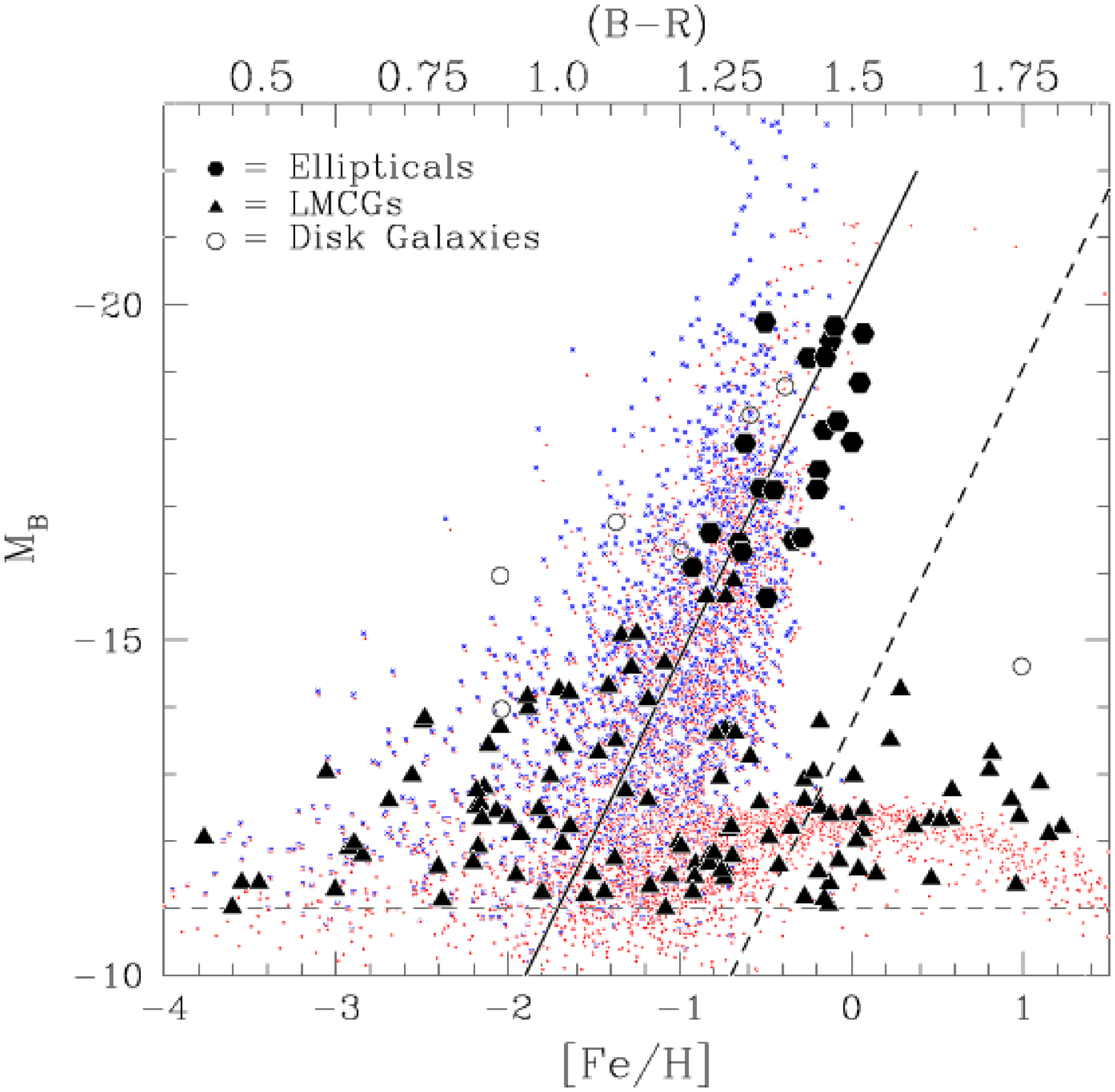}}
\end{center}
\figcaption{The (B-R) color-magnitude diagram, plotted also in terms
of magnitude-metallicity, for simulated and actual data from Abell~0426.  
 The large (black) solid symbols are the 
Abell~0426 cluster data from Conselice et al. (2002a,b). The solid line shows
the color-magnitude relationship for the giant ellipticals and the dashed
line denotes the 2$\sigma$ cut between LMCGs considered blue $<2\sigma$ 
and red $>2\sigma$.  The small (blue) crosses 
are the $\lambda$CDM modeled galaxies from Nagamine et al. (2001). The small
(red) points are the location of these galaxies after they have gone through 
the tidal stripping simulation discussed in this letter.}
\end{inlinefigure}

Conselice et al. (2002b) suggest that the red LMCGs galaxies are 
cluster members that did not originated within the cluster, but are the 
remnants of
stripped down larger galaxies that were accreted into the cluster in the 
last few Gyrs.  As such, these red LMCGs could
be acting as masquerading dwarf spheroidals that were not part of the initial
mass function of the cluster as predicted in numerical and semi-analytic
simulations (e.g., Springel et al. 2001; Nagamine et al. 2000b).
How do these red LMCGs affect the computed Abell~0426 luminosity function?
The total Abell~0426 cluster luminosity function is shown in Figure~2
as solid circles and as a solid fitted line.   
The luminosity function of the red and blue LMCGs 
are plotted as open and solid triangles and as long-dashed and dotted lines 
in Figure~2.  These luminosity functions are complete down to our 
limit of $M_{B} = -11$ as discussed in Conselice et al. (2002a).    
The faint-end of these luminosity functions are fit by the Schechter function 
and a power-law.  The Schechter
function fit gives the parameters $\phi^{*}, L^{*}$ and $\alpha$; which are
the normalization, characteristic luminosity, and faint-end slope.  
While the power law form is given by $dN/dL = \phi^{*} L^{\alpha}$.

The resulting faint-end computations of $\alpha$ are found
to be consistent between 
these two fitting methods. For the entire Abell~0426 galaxy population, we find
$\alpha$ = -1.44$\pm$0.04 (solid line).  When we consider just
the blue population, $\alpha$ = -1.25$\pm$0.05 (dotted line), which is 
different from the total Abell~0426 cluster luminosity
function at a significance of $> 3.5 \sigma$. For the red LMCGs
a power-law fit gives $\alpha$ = -1.82$\pm$0.05 (long dashed line), which is 
$> 10\sigma$ different from the blue luminosity function.   For comparison 
purposes we calculate the luminosity function (short dashed line) 
of the objects we identify as background systems (crosses), finding  
$\alpha$ = -1.81$\pm$0.03.

This suggests that there is a strong color
dependence on the luminosity function of Abell~0426 such that the LF becomes
steeper at the faint end due to the presence of red LMCGs.  
We are not the first to identify steepening LFs due to different
types of low-mass galaxies in clusters, however.  Boyce et al. (2001) argue 
from observations of a $z = 0.154$ cluster that its luminosity function is
steepened by a population of blue starbursting dwarfs. 

The $\alpha$ = -1.25 slope has been considered the `universal'
luminosity function faint end slope since Schechter (1976) and is
the slope of the field luminosity function computed using
Sloan Digitized Sky Survey (SDSS) data (Blanton et al. 2001;
cf. Cross et al. 2001 for the 2dF LF giving $\alpha$ $\sim$ -1.1).  Clusters 
have fitted $\alpha$ values that range from -0.9 to -2.2 depending upon
the cluster, area sampled, and method of computation 
(e.g., Trentham 1998).  However, semi-analytic and numerical models
predict LFs that are generally flatter than $\alpha = -1.25$ (Kauffmann,
Nusser \& Steinmetz 1997; Springel et al. 2001; Nagamine et al. 2001b; 
Benson et al. 2001; although
cf. Kauffmann et al. 1997, $\alpha$ = -2 for dark matter halos). As the
LF slope for the blue LMCGs in Abell~0426 are near the field values 
and close to what models predict, the intriguing possibility exists that 
there is a universal initial
galaxy mass function that is altered by cluster dynamical effects that
create these red LMCGs.

\section{A Universal Initial Galaxy Mass Function?}

Is there an universal initial mass function for galaxies that is independent
of environment?   Could the large differences in luminosity functions
observed in clusters and in the field (e.g., Binggeli et al. 1988) 
be the result of evolution induced by dynamical effects acting on a universal 
luminosity function? 
To explore the idea that cluster processes could change the luminosity
functions of clusters, we simulate the evolution of a cluster's LF as it
undergoes a simple analytic dynamical stripping processes.   This is done by 
using the field luminosity function of $>11,000$ galaxies from the SDSS 
(Blanton et al. 2001).  We convert the SDSS g${^*}$ band LF into a B-band 
LF by using the conversion factor given in Fukugita et al. (1995).  The
resulting Schechter function is given by the parameters, $\phi$ = 
0.007~Mpc$^{-3}$, M$_{*} = -21.36$ and $\alpha = -1.26$. 

We assume that the SDSS field LF is indicative of the passive evolution of
the galaxy mass function at $z \sim 0$.  This assumption has some basis in 
actuality as interactions between field galaxies are not
common, as they are by definition in low-density environments.
To simulate the formation of a cluster LF, we assume
that the initial Abell~0426 LF resembled the Blanton et al. (2001) field 
LF at z = 0, except
that $\phi$ is normalized to match the Abell~0426 galaxy density.  
We then simulate the effects of tidal 
interactions on these galaxies and compute the resulting LF. After
performing this simulation, we find that the 
faint-end slope, $\alpha$, becomes steeper.

\begin{inlinefigure}
\begin{center}
\resizebox{\textwidth}{!}{\includegraphics{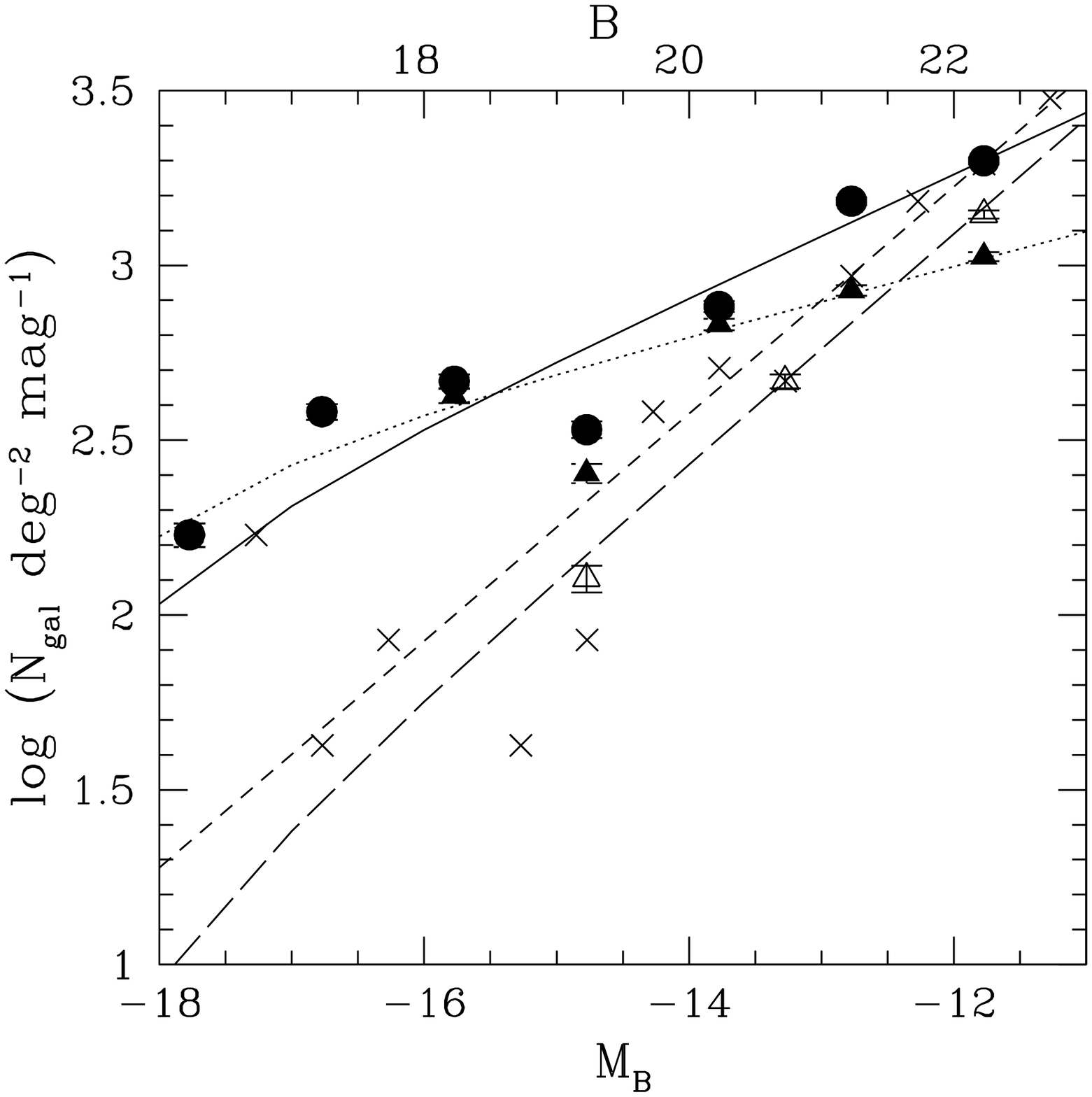}}
\end{center}
\figcaption{The Abell~0426 cluster luminosity function plotted in
various color selected methods down to our completeness
limit of $M_{B} = -11$.  The solid round points and solid
fitted line are the total luminosity function of the central region of
Abell~0426 (Conselice et al. 2002a).  The luminosity function of the red and 
blue LMCGs are plotted as open and solid triangles and long dashed and
dotted lines.  The crosses mark the density of background galaxies at each
magnitude bin whose luminosity function is fit as the short dashed line.}
\end{inlinefigure}

We demonstrate this through limited Monte Carlo simulations of mass-loss 
induced through high-speed impulse galaxy
interactions.   The model we use follows the general analytic procedure 
described in Aguilar and White (1985) and Conselice et al. (2002b).  These
analytic computations are for simple spherical galaxies,
which is an over simplification to most cluster galaxies.  The
dynamical stripping process is furthermore generally more complicated than 
the simple model we use, as the physics involves
several components we are not considering, including: the distribution of mass
in individual galaxies, the clumpiness of the cluster mass
distributions (Gnedin 1999), and the
closest approach towards the center of the cluster (Merritt 1994).  With
these caveats in mind, we model the mass evolution of galaxies in
the central
300 kpc area of a rich cluster with properties similar to Abell~0426. 
Given these parameters, most galaxies should pass well 
within the center of the cluster and tidal effects should be quite
pronounced (e.g., Merritt 1984; Gnedin 1999).  Using these assumptions 
the total initial mass M$_{0}$ is related to the final total 
mass, M$_{f}$, after a single
interaction by: M$_{f} =$ M$_{0}$$(1-\eta_{t})$ 
with a total mass loss rate $\delta$M/M $= \eta_{t}$ 
(Conselice et al. 2002b) per interaction.

We simulate the stripping process by computing the evolutionary 
history of 
each galaxy from the SDSS field LF after it goes through the Monte Carlo
simulation.  The parameters varied are the stellar mass-loss
rate, $\eta_{*}$, and the number of interactions which is chosen from 
a distribution function having a maximum number of possible interactions,
q.    The distribution of the number of interactions
galaxies can undergo in the simulation is divided into two equal components.
One half of the galaxies undergo a low number of interactions equally 
distributed from 1 to 10.  
The other half of the galaxies all undergo q interactions, 
which is changed for each simulation.  
To determine the stellar mass loss rate we take into account
the effects of different dark matter distributions in galaxies by
including a factor that corresponds to the fact that low-mass galaxies
are dark matter dominated (Peterson \& Caldwell 1993).  Because 
low-mass systems have higher M$_{\rm total}$/L
ratios, and possibly extended dark halos, it is therefore harder to remove
stellar mass located at their centers.  To account for this we
assume that the stellar mass loss rate, $\eta_{*}$, is inversely proportional 
to 
the M$_{\rm total}$/L ratio, $\eta_{*} 
\sim \eta_{t}$(M$_{\rm total}$/L)$^{-1}$ $\sim
\eta_{t}$(L$^{0.4}$) (e.g., Dekel \& 
Silk 1985), which we arbitrarily normalize at M$_{B} = -20$.  Note that the
Aguilar \& White (1985) mass loss rates we use are independent of the
total mass or M$_{\rm total}$/L ratio of a galaxy, but as we explicitly 
assume that lower mass cluster 
galaxies contain extended dark matter halos, the stellar mass loss rate,
$\eta_{*}$,
is a function of mass and M$_{\rm total}$/L.  Otherwise, if we do not include
this dependence on the M$_{\rm total}$/L ratio, low-mass systems are
evaporated quickly in the simulation and the LF never becomes
much steeper than the universal value.

Through numerical
and analytical models, a reasonable maximum
mass loss rate per interaction is $\eta_{t} \sim 0.1$ for a relatively strong 
interaction using the conditions in Abell~0426 (e.g., Aguilar \& White 1985; 
Conselice et al. 2002b).
Our simulations therefore select a random $\eta_{t}$ value from 0 to
0.1 for each interaction.  The number of interactions the galaxies in
the simulations undergo is given by the distribution function described 
above.   After a simulation is run, the result is a new galaxy luminosity
distribution from which a luminosity function is computed and then fit 
after normalizing the number of galaxies produced down to M$_{B} = -11$
equal to the number observed in the center of Abell~0426 as
discussed in Conselice et al. (2002a). 

How does the faint end of the LF change as a function of the maximum number of 
interactions, q?  This is
shown in Figure~3.  The faint-end slope $\alpha$ decreases until
the number of interactions reaches $\sim$400 where $\alpha \sim -2$.
The efficiency of the mass-loss processes becomes high enough such
that at q$>400$ a significant number of galaxies become fainter
than the magnitude limit M$_{B} = -11$.  This is a reasonable number of
interactions in a rich cluster as high-speed encounters are effective out to 
many times the scale lengths of typical galaxies (e.g., Aguilar \& White 
1985), and
there are $\sim200$ galaxies in the inner 300 kpc of Abell~0426, which a 
cluster galaxy will transverse every 0.3 Gyrs.

This simple model demonstrates that by considering the evolution of initial
luminosity functions through mass-loss induced by high speed encounters
between galaxies the luminosity function becomes steeper.  These models
also suggest that the observed LF down to M$_{B} = -11$ becomes flatter in 
denser and older areas 
where more interactions have occurred. This is consistent with finding  
flatter LFs in the central parts of clusters, as these
are the densest and oldest locations and thus areas where stripping
processes can create a deficiency of low-mass systems (see also 
Oemler 1974; Lopez-Cruz et al. 1997; Adami et al. 2000).  

\section{Red Low-Mass Cluster Galaxies}

We briefly re-address in this section the issue of red faint cluster
galaxies recently found in a host of nearby clusters (e.g., Adami
et al. 2000; Rakos et al. 2001; Mobasher et al. 2001; Conselice et al. 2002a,b)
to show that these galaxies could be the remnants of tidally
stripped higher mass systems.  It is clear that some tidal
stripping has occurred in the past due to the presence of a significant
amount of intracluster light and tidal debris tails (Gregg \& West 1998) as
well as from examples of cluster galaxies that appear to be in the process of
undergoing disruption, but with no obvious nearby neighbors (Conselice \& 
Gallagher 1998, 1999).

\begin{inlinefigure}
\begin{center}
\resizebox{\textwidth}{!}{\includegraphics{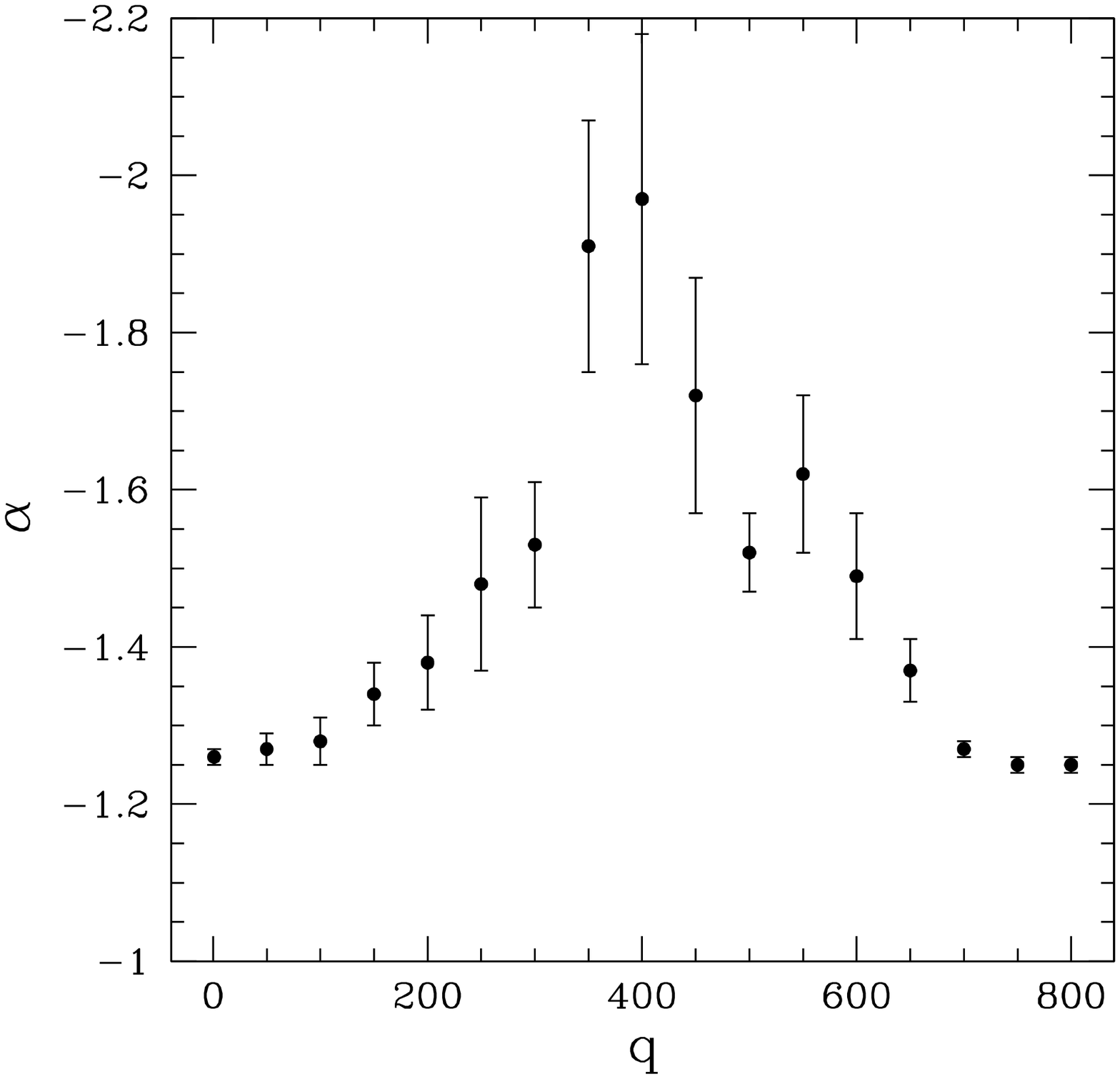}}
\end{center}
\figcaption{The cluster luminosity function
slope, $\alpha$, as a function of the number of maximum interactions each
galaxy undergoes during evolution in a Abell~0426 like cluster.}
\end{inlinefigure}

We examine the resulting metallicity, and inferred color, distributions of 
galaxies after they undergo the stripping process through
the use of $\lambda$CDM modeled galaxies produced by Nagamine et al.
(2001a).  Nagamine et al. (2001a) produce through hydrodynamic simulations
galaxies with various metallicities, luminosities and star-formation 
histories.  We use these galaxies as the initial cluster sample.  
Each of these galaxies is sent through the Monte Carlo simulation, described
in \S 3,
to determine the amount of mass-loss occurring after q=300 interactions when
$\alpha \sim -1.5$ (Figure~3). Figure~1 shows the resulting 
magnitude-metallicity-color diagram where
the small crosses are the original $\lambda$CDM galaxies and the dots show
where this same population falls on this diagram after they go through the
simulation described in \S 3.

We implement a color evolution in these systems by
assuming these galaxies follow the average (B-R) color gradients seen
in spiral bulges (Balcells \& Peletier 1994) and by using the trend between
the radii and magnitudes of galaxies as described in Binggeli \&
Cameron (1991). After mass stripping occurs in a galaxy it
will become fainter, and smaller, due to the strong correlation between galaxy
size and magnitude (Binggeli \& Cameron 1991).  Since the inner parts of
galaxies are redder, and metal rich (e.g., Balcells \& Peletier 1994), 
the stripping process produces fainter, more metal-rich, and hence
redder, galaxies.   
The metallicity [Fe/H] of the Nagamine et al. simulated galaxies are converted
into a (B-R) color
using an empirical relationship between these two parameters found by Harris 
(1996) for old globular clusters.  

LMCGs and giant ellipticals in Abell~0426, as discussed in Conselice et al. 
(2002a,b), are plotted on Figure~1.  
The new magnitude-metallicity diagram (dots) changes from the 
initial one (crosses) due to the effects of mass-loss, and as a result a 
population of 
red low-mass cluster galaxies is created.    The more luminous, and thus
more massive, systems lose more stellar mass than the fainter systems
and hence their remnants are faint and red.  The lower mass systems, because
of their higher M$_{\rm total}$/L ratios, lose a lower fraction of their 
stellar light and thus become slightly fainter and more metal-rich, but not as
much as the higher luminosity systems do.  The result of this
is the creation of a remnant population with different colors but with 
similar luminosities, within the range M$_{B}$ = -10.5 to -12.5.
This simple example shows that the high-speed interaction process could
be responsible for producing the observed number of red low-luminosity cluster
galaxies that are `artificially' steepening the luminosity functions of
clusters.   This also implies that tidal stripping of material from these
galaxies is at least partially the cause of intracluster light and cD halos
(Gallagher \& Ostriker 1972).

The simulations and results presented in this letter are more general than 
that implied by the assumptions of the underlying physical effects discussed.
Two things are firmly suggested by our results. The first is that no matter 
what the origin of the objects identified as red LMCGs are, they are steepening
the LF, and future LF determinations should account for this effect.  
The second is that any cluster stripping process that removes
mass, including several alternative scenarios we have not considered, will 
naturally produced galaxies with low-masses and red colors.

I thank Jay Gallagher and Rosie Wyse for their participation in the
survey of Abell~0426 and K. Nagamine for kindly supplying his $\lambda$CDM
model results used in this letter. I also acknowledge useful conversations
with Andrew Benson on some of the topics discussed in this letter.  I also 
thank Claudia Scarlata and the referee for very useful comments which 
improved the presentation of this letter. This work was supported in part by 
NSF grants AST-9803018 and AST-9804706.

\end{document}